\documentclass[12pt]{iopart}
\usepackage{graphicx}
\usepackage[utf8]{inputenc} 
\usepackage[T1]{fontenc}
\usepackage{color}
\usepackage{soul} 
\usepackage[
,textwidth=15cm
,textheight=20cm
,verbose
,dvips
]{geometry}
\usepackage[numbers,comma,square,sort&compress,merge]{natbib}

\graphicspath{{figs/}}

\begin{document}

\title[Top-down fabrication of ultrathin GaN nanowires]{A route for the top-down fabrication of ordered ultrathin GaN nanowires}
%should we change the title mentioning the mechanical instability of the NWs?

\author{M.\,Oliva, V.\,Kaganer, M.\,Pudelski, S.\,Meister, A.\,Tahraoui, L.\,Geelhaar, O.\,Brandt, T.\,Auzelle}

\address{Paul-Drude-Institut für Festkörperelektronik, Leibniz-Institut im Forschungsverbund Berlin e.V., Hausvogteiplatz 5–7, 10117 Berlin, Germany}

%\address{Current address: NOW GmbH, Fasanenstraße 5, 10623 Berlin}

%\author{M.\,Oliva$^1$, V.\,Kaganer$^1$, M.\,Pudelski$^1$, S.\,Meister, A.\,Tahraoui, L.\,Geelhaar, O.\,Brandt, T.\,Auzelle}
%
%\address{$^1$Paul-Drude-Institut für Festkörperelektronik, Leibniz-Institut im Forschungsverbund Berlin e.V., Hausvogteiplatz 5–7, 10117 Berlin, Germany}
%
%\address{$^2$Current address: NOW GmbH, Fasanenstraße 5, 10623 Berlin}

\ead{oliva@pdi-berlin.de}
\ead{auzelle@pdi-berlin.de}
\vspace{10pt}

\begin{abstract}
Ultrathin GaN nanowires (NWs) are attractive to maximize surface effects and as building block in high-frequency transistors. 
Here, we introduce a facile route for the top-down fabrication of ordered arrays of GaN NWs with aspect ratios exceeding $10$ and diameters below $20\,$nm. Highly uniform thin GaN NWs are first obtained by using electron beam lithography to pattern a Ni/SiN$_x$ hard mask,
%by producing a hard mask of Ni/SiN$_x$ by using electron beam lithography 
%are first obtained by electron beam lithography patterning of a Ni/SiN$_x$ hard mask 
followed by dry etching and wet etching in hot KOH. The SiN$_x$ 
is found to work as an etch stop during wet etching in hot KOH. Arrays with NW diameters down to $(33 \pm5)$\,nm can be achieved with a yield exceeding $99.9\,\%$. Further reduction of the NW diameter down to $5$\,nm is obtained by applying digital etching which consists in plasma oxidation followed by wet etching in hot KOH. The NW radial etching depth is tuned by varying the RF power during plasma oxidation. NW breaking or bundling is observed for diameters below $\approx 20$\,nm, an effect that is associated to capillary forces acting on the NWs during sample drying in air. This effect can be principally mitigated using critical point dryers. Interestingly, this mechanical instability of the NWs is found to occur at much smaller aspect ratios than what is predicted for models dealing with macroscopic elastic rods. Explicit calculations of buckling states show an improved agreement when considering an inclined water surface, as can be expected if water assembles into droplets. The proposed fabrication route can be principally applied to any GaN/SiN$_{x}$ nanostructures and allows regrowth after removal of the SiN$_{x}$ mask. 
%below 300 words - check
\end{abstract}

\vspace{2pc}
\submitto{\NT}
%try to publish in the NW special issue
%https://publishingsupport.iopscience.iop.org/journals/nanotechnology/
\maketitle
\ioptwocol

\section{Introduction}
Downscaling of Si-based devices was shown to be decisive to increase performance, expand functionalities and decrease production costs \cite{Chen_2021}. Along the same line, miniaturization of GaN three-dimensional structures like wires and fins has today become relevant for both electronic and photonic applications \cite{Shinohara_2013,Jo_2015,Fatahilah_2019a,Wasisto_2019,Meneghini_2021,Mikulics_2022}. 
%\hl{add more citations? see commented text}
%examples of GaN based FETs\cite{Jo_2015,Yu_2016,Hu_2017a,Son_2019,Doundoulakis_2019,Fatahilah_2019a} and LEDs\cite{Bavencove_2010,Bavencove_2011,Tchernycheva_2014,Guan_2016,Hartensveld_2019,Wasisto_2019,Yulianto_2021,Mikulics_2022,Pandey_2022,Yanagihara_2022}
Specifically, ultrathin GaN nanowires (NWs) with diameters $\leq 20$\,nm are beneficial for achieving ultrafast switching in field-effect transistors \cite{Chowdhury_2017,Thakur_2021}, they can elastically relax large amounts of epitaxial strain \cite{Glas_2006,Oto_2019,Teng_2016}, 
%\hl{I don't like the citation of Avit. I would remove it and add instead:Niquet 1997,Wolz 2013,Zhang 2014a,Teng 2016,Kaganer 2016. Also, note that these references are about thin NWs, not about ultrathin NWs.}
and they host dielectrically confined excitons up to room temperature \cite{Zettler_2016}. If arranged deterministically on the substrate surface, such NW arrays can principally form photonic cavities \cite{Oto_2021} or 
%electrical 
interconnects \cite{Car_2014}.

%Ultrathin GaN NWs with diameters down to $8$\,nm can be directly fabricated top-down by self-assembling a Si mask on GaN(0001) followed by thermal sublimation of GaN \cite{Damilano_2016,Damilano_2017}. Thin NWs with diameters down to $15\,$nm can be also grown in a self-assembled bottom-up fashion\cite{Stoica_2008}.
%However, a deterministic patterning of the GaN(0001) surface as provided by electron beam lithography (EBL) is desired to gain control over the NW position and dimensions. 
%%So far, EBL patterning of SiN$_x$, SiO$_x$, Ni or Ti masks deposited on GaN has typically resulted in nanostructures at least 25\,nm wide \cite{Barbagini_2011,Kano_2015}. 
%%In contrast, ordered arrays of GaN NWs
%%It follows that 
%So far, GaN NWs produced bottom-up (top-down) from patterned masks feature, neglecting tapering, a minimum diameter of $25$\,nm \cite{Kano_2015} ($40$\,nm \cite{Teng_2016}). 
Using self-assembled processes, randomly placed NWs with ultrathin diameters can readily be obtained in a top-down or bottom-up fashion, with a minimum diameter of $8$ \cite{Damilano_2016,Damilano_2017} and $15$\,nm \cite{Stoica_2008}, respectively. However, if the substrate surface is deterministically patterned---\textit{e.g.} using electron beam lithography (EBL)---to gain control over the NW position and dimensions, diameters larger than $25\,$nm are systematically obtained if neglecting tapering \cite{Kano_2015,Teng_2016}. 
Ordered arrays of thinner GaN NWs could still be achieved through additional processing steps. Thermal sublimation in vacuum or NH$_3$ environment allows reducing the NW diameter down to a few nanometers \cite{Brockway_2011a,Zettler_2016}. Yet, the thinning rate depends exponentially on temperature, a parameter that is in practice difficult to change abruptly in vacuum chambers. Direct wet etching in solutions of tetramethylammonium hydroxide can reduce NW diameter, but the etching rate is strongly anisotropic \cite{Jo_2015,Im_2016,AlTaradeh_2021}. 

%Ultrathin GaN NWs can be directly fabricated top-down by self-assembling a Si mask on GaN(0001) followed by thermal sublimation of GaN \cite{Damilano_2016}. The obtained NWs are randomly placed on the sample surface and typically feature diameters of ($8 \pm 3$)\,nm and lengths of $\approx100$\,nm \cite{Damilano_2017}. However, a deterministic patterning of the GaN(0001) surface as provided by electron beam lithography (EBL) is desired to gain control on the NW position and dimensions. So far, EBL patterning of SiN$_x$, SiO$_x$, Ni or Ti masks deposited on GaN has typically resulted in nanostructures at least 25\,nm wide \cite{Barbagini_2011,Kano_2015}. It follows that GaN NWs produced from such patterned masks feature a minimum diameter of $25$\,nm for selective area growth \cite{Kano_2015} and of $40$\,nm for selective area etching (neglecting tapering) \cite{Teng_2016}. Thinner GaN NWs can still be achieved through additional processing steps. Thermal sublimation in vacuum or NH$_3$ environment allows reducing NW diameter down to a few nanometers \cite{Brockway_2011a,Zettler_2016}. Yet, the thinning rate depends exponentially on temperature, a parameter that is in practice difficult to change abruptly in vacuum chambers. Direct wet etching in solutions of tetramethylammonium hydroxide can reduce NW diameter, but the etching rate is strongly anisotropic \cite{Jo_2015,Im_2016,AlTaradeh_2021}.
More attractive for a nanometer-scale etching is the use of digital etching or atomic layer etching for which the etching depth is controlled by the number of applied etching cycles \cite{Keogh_2006,Burnham_2010,Liu_2015a,Sokolovskij_2016a,Gao_2018,Chiu_2018,Wu_2019,Johnson_2019,Ruel_2021,Hwang_2021,Shih_2022}. So far, most of these processes have been developed on polar GaN and (Al,Ga)N surfaces for recess etching of III-nitride transistors. Only one recent work deals with semipolar facets \cite{Shih_2022} but little is known about digital etching of nonpolar facets corresponding to the sidewalls of GaN NWs.

Here, we demonstrate a facile route for top-down fabrication of ordered arrays of vertical GaN NWs with diameters $\leq20$\,nm and aspect ratios larger than $10$. Highly uniform arrays of GaN NWs with a minimum diameter of $(33 \pm5)$\,nm are first obtained by using EBL to pattern a Ni/SiN$_x$ hard mask
%by producing a hard mask of Ni/SiN$_x$ by using EBL 
%are first obtained by electron beam lithography patterning of a Ni/SiN$_x$ hard mask 
%produced by EBL patterning of a Ni/SiN$_x$ mask 
followed by a combination of dry and wet etching. The SiN$_x$ layer is found to act as an etch stop during wet etching in hot KOH. Further reduction of the NW diameter down to $5$\,nm is obtained by applying digital etching consisting in plasma oxidation and wet etching in hot KOH. However, below $20$\,nm in diameter, NWs with large aspect ratio are seen to collapse as a result of capillary forces acting during sample drying. The early collapsing of the NWs cannot be accounted for by models developed so far for macroscopic elastic rods. Instead, our explicit calculations of the NW bent states under capillary forces show that a critical parameter governing the NW mechanical instability is the inclination angle of the water surface with respect to the NW axis.

\section{Experiments}

\subsection{Top-down NW fabrication}
\label{Sec:Methods_1}

%The best-established top-down fabrication technique to obtain exceedingly homogeneous arrays of thin NWs for GaN relies on inductively coupled plasma (ICP) etching followed by an anisotropic wet etching step with a KOH solution to remove the dry-etch damage\cite{Li_2011,Li_2012}. The ICP etching has been shown to introduce lattice defects\cite{Cao_1999} and deep traps\cite{Cho_2008}, which dramatically reduces the PL intensity\cite{Saotome_1996,Choi_2000,Keller_2006} and compromises device perfomance\cite{Cao_2000,Cao_2000a}. Since $1996$ it is known that this damaged material can be removed by KOH etching\cite{Saotome_1996}, NaOH\cite{Cao_2000}, HF followed by rapid thermal annealing (RTA)\cite{Choi_2000}, annealing in an RTA\cite{Cao_2000a} or in a Metalorganic vapour-phase epitaxy (MOVPE) reactor\cite{Keller_2006}. Li \textit{et al}\cite{Li_2011} where the first ones applying this concept to NWs with sub wavelength diameter in $2011$.

To produce the initial GaN NW array with high uniformity, we follow a conventional top-down fabrication scheme comprising mask patterning, plasma etching and wet etching in concentrated KOH. This approach schematized in Fig.\,\ref{01_synthesis}(a) was shown effective for synthesizing micrometer long GaN NWs with a minimum diameter of $\approx100$\,nm that remained optically active in spite of the plasma treatment \cite{Li_2011,Li_2012}. To achieve smaller NW diameters, we rely here on EBL for mask patterning.

The substrates are commercial GaN layers grown epitaxially on sapphire with estimated donor concentrations of $10^{16}$\,cm$^{-3}$ [non intentionally doped (\emph{nid})-GaN] and $10^{18}$\,cm$^{-3}$ (\emph{n}-GaN). The chosen mask consists in a vertical stack of SiN$_x$ and Ni. A $20$\,nm thick SiN$_x$ layer is first deposited by magnetron radio-frequency (RF) sputtering (PRO line PVD$75$ from Kurt J. Lesker Company) at a power of $160$\,W, a pressure of $1.9$\,mTorr and an Ar:N ratio of $1$. The layer is further treated by rapid thermal annealing at $600\,^{\circ}$C for $10\,$min. Next, an adhesion layer AR300-80 is spin-coated and baked for $2$\,min at $180\,^\circ{}$C. It is covered by an $80\,$nm thick CSAR $62$ positive resist (Allresist GmbH) deposited by spin-coating, baked for $1$\,min at $180\,^\circ{}$C and eventually exposed by EBL (RAITH150 Two EBL). The printed pattern is made of circular patches arranged in a hexagonal symmetry within a $50 \times 50\,$µm$^2$ quadratic microfield. Various microfields are printed with pitches ranging from $0.2\,$ to $15\,$\textmu{}m (nearest-neighbor center-to-center distance), and patch diameters ranging from $30$ to $120\,$nm. The electron dose is varied from $270$\,\textmu{}C/cm$^2$ for the smaller features to $140$\,\textmu{}C/cm$^2$ for the bigger ones. The pattern is developed by successive baths in developer AR600-546 and stopper AR600-60 solutions for $70$ and $60$\,s, respectively, leaving holes in the resist layer. 
%\hl{Then, an O$_2$ plasma step at $0.5\,$mBar for $12\,$s in a Diener electronic plasma reactor is performed.}
The sample is then coated with a $25\,$nm thick Ni layer by metal evaporation (Pfeiffer classic $500$) and mask lift-off is eventually performed with the remover AR $600$-$71$ for $60\,$min. The obtained Ni pattern will be used as a hard mask for the subsequent plasma etching. The non-protected part of the SiN$_x$ layer is first reactively etched in CF$_4$/O$_2$ at a plasma power of $50\,$W and pressure of $4\,$Pa (Sentech SI $500$). Next, GaN is anisotropically etched via inductively coupled plasma reactive ion etching (ICP) at a pressure of $1.3\,$Pa with a BCl$_3$:Cl$_2$ gas ratio of $20$:$5$, an RF power of $25\,$W and an ICP power of $100\,$W (SAMCO RIE-$140$iP). Ni is then removed in a ferric chloride solution for $1\,$min at $40$--$60\,^{\circ}$C (Ni etchant type I from Transene).
Finally, the formed GaN NWs are dipped for $70$--$130\,$s in either a $6.9\,$M\,KOH aqueous solution or in the AZ$400$K developer which also contains KOH. The temperature of the KOH solution ranges between $40$ and $80\,^{\circ}$C, as measured by a thermometer placed in the beaker.

%the investigated samples are pieces from GaN21, GaN15 and GaN12 (see folder 'nanowires\Data_Experiments\Series+Projects\GaN\2020-2021_Pudelski_NWThinning\Samples\'). Samples GaN12, GaN15 and GaN21 were annealed by RTA at 600deg for 10min

\subsection{Digital etching}

The digital etching examined here consists in two steps per cycle and is applied on GaN NWs with their top facet still capped by SiN$_x$. The NWs are first oxidized in an O$_2$ plasma for $15$\,min (SAMCO RIE-140iP). The O$_2$ plasma is at a pressure of $7$\,Pa ($10$\,sccm of O$_2$) with an ICP power of $100$\,W and an RF power of $20$--$50$\,W. A supplementary NW sample was treated with an ICP power of $50$\,W and an RF power of $15$\,W. The oxidized NWs are eventually dipped in a $6.9\,$M\,KOH aqueous solution at $60\,^\circ{}$C for up to $5$\,min.

\section{Results}

%\subsection{Fabrication of thin GaN NWs}
%\subsection{Ordered, top-down, thin GaN NWs}
\subsection{NW array morphology}
%Part I: how to fabricate the thin NWs, discuss control of diameter, length, and stability of wet etching process, first observations/results 

%=====================================================================
%%Fig.1
\begin{figure*}
	\includegraphics*[width=\textwidth]{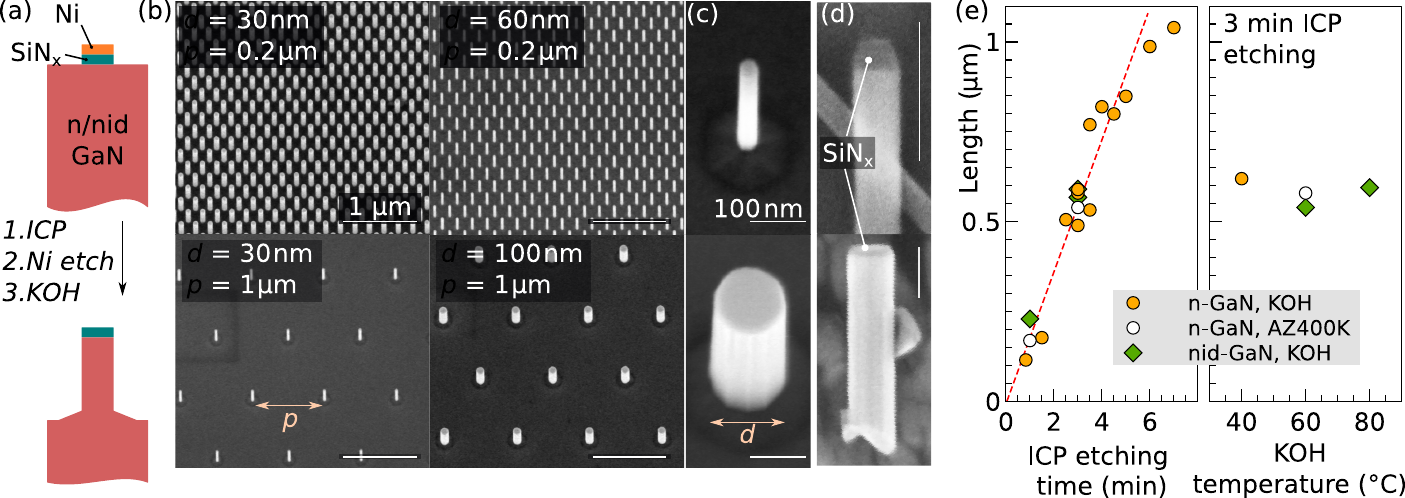}
	\caption{\label{01_synthesis}(a) Main processing steps used to achieve the top-down synthesis of thin GaN NWs. (b)--(c) Representative bird’s-eye view ($15\,^{\circ}$ tilt from top-view) secondary electron micrographs of the obtained GaN NWs. Nominal values for the diameter $d$ and pitch $p$ of the initial mask are given in (b). A magnified view of single NWs is shown in (c). (d) Cross section secondary electron micrographs of a thin and a thick dispersed NW, featuring their $20\,$nm-thick SiN$_x$ top patches. (e) Averaged lengths of the obtained NWs as function of ICP etching time and temperature of the KOH treatment. The initial GaN template are either \emph{n}-doped (\emph{n}-GaN) or non intentionally doped (\emph{nid}-GaN).
	}
%maybe shift images in (d) to figure 2?
%%b,c: MIO_GaN#6A
%%d, top: MIO_GaN15A, microfield 40_200
%%d, bottom: MIO_GaN6A, microfield 80_200
%%e, left: MIO_GaN#20C, MIO_GaN#4H,F, MIO_GaN#20E, MIO_GaN#4I, MIO_GaN#5C,D, MIO_GaN#6A,B, MIO_GaN#12C,D,G, MIO_GaN#15A,B,C (SiN+Ni mask)
%%not shown, but described in text in context of d: MIO_GaN#15D,G (SiN mask)
%%d,right: MIO_GaN#4I, MIO_GaN#5D, MIO_GaN#6A, MIO_GaN#6B
\end{figure*}

% Add Motivation?
%qualitative and general description of the NW morphology
Arrays of vertical GaN NWs produced here feature a yield of $99.9\,$\% as can be seen in the representative secondary electron micrographs of Fig.\,\ref{01_synthesis}(b). For these NWs, ICP etching was carried out for $3\,$min and KOH treatment was $130$\,s long at $60$\,$^{\circ}$C in aqueous solution. The $0.6$\,\textmu{}m long NWs are all vertical and exhibit excellent uniformity. The magnified views of single NWs shown in Fig.\,\ref{01_synthesis}(c) reveal a constant diameter along the NW length and the presence of smooth sidewalls. These well-defined sidewalls result from the final KOH treatment, which is reported to remove defects induced by the ICP etching \cite{Li_2012,Debnath_2014,Liu_2015}.  

%discuss SixNy patch -> diameter discussion
Cross-sectional secondary electron micrographs views of randomly dispersed NWs [Fig.\,\ref{01_synthesis}(d)] allow visualization of the SiN$_x$ top patch. It has a comparable thickness as the initial sputtered SiN$_x$ layer and features a tapered morphology, likely resulting from the ICP etching. Importantly, the base of the SiN$_x$ patch has a similar diameter as the GaN NW underneath, suggesting that both are correlated. 
%defines the GaN NW thickness. 
%The diameters of the GaN NW and of the Si$_x$N$_y$ patch are equal for all NWs. 
%add discussion/comparison with literature?

%length control (add a motivation?)
Figure \ref{01_synthesis}(e) shows the average length of NWs obtained for different ICP etching times. Regardless of the donor concentration in the GaN template and the type of KOH solution, the graph evidences a linear relationship up to etching times of $6$\,min. For longer etching time, the NW length saturates at about $1.1$\,\textmu{}m and the yield is drastically reduced, which is associated with a complete consumption of the $45$\,nm thick Ni/SiN$_x$ hard mask. The ratio between the GaN and the Ni/SiN$_x$ etching rates indicates an etching selectivity of $\approx 24$, a value that compares favorably with prior reports \cite{Chang_2001,Okada_2017,Ahmad_2021}. In the absence of Ni, NWs only $0.1$\,\textmu{}m long are obtained, which translates in an etching selectivity between GaN and SiN$_x$ of $5$, similar to Ref.\,\cite{Okada_2017}. The best reported selectivity between GaN and Ni amounts to $\approx~20$ \cite{Ahmad_2021}. By taking the selectivity of individual SiN$_x$ and Ni layers, our hard mask should thus provide a maximum NW length of only $0.6$\,\textmu{}m. Our enhanced selectivity compared to Ref.~\citenum{Ahmad_2021} may be due to an improved adhesion of the mask provided by the underlaying SiN$_x$ layer and/or to less aggressive etching conditions.
Figure\,\ref{01_synthesis}(e) also evidences that the final NW length has no marked dependence on the temperature of the KOH bath, on the nature of the KOH solutions and on the doping level of the initial GaN layer. Similar conclusions are drawn concerning the NW diameter. Interestingly, straight NW sidewalls are already obtained after $130\,$s, whereas etching times of several hours were required in Refs.\,\cite{Li_2011,Debnath_2014}. We associate the faster etching observed in this work to the smaller volume of material to be etched and to the higher temperature of the KOH bath. 

% A similar conclusion holds for the NW diameter. We additionally find that the diameter of the NWs is unaffected by the different KOH etching recipes. These results demonstrate that the KOH-based wet etching is fast and robust, and suggest that the process not only removes the damage layer created by dry etching\cite{Saotome_1996,Li_2011}, but it is also limited by the wet etching mask's thickness.
% %because of the 
% %\hl{Thomas: I disagree with the statement that KOH is removing damaged GaN. From our data, I believe that the KOH etches everything that is not capped by SiN (or Ni).}
% %MO: I am not sure what it means if we claing that the ICP/KOH process is self limited
% 
% %effect of KOH: compare with other studies
% Other studies on top-down GaN NWs also reported that the NW length remains invariant after KOH etching sessions of different times\cite{Li_2011,Debnath_2014}. Li \textit{et al}\cite{Li_2011} found that perfectly straight NWs can be achieved only after $6\,$h of KOH etching at room temperature. Increasing the KOH etching temperature to $80^{\circ}$C slightly speeds up the process to $2\,$h, as reported by Debnath \textit{et al}\cite{Debnath_2014}. In contrast, we obtain perfectly straight GaN NWs with smooth non polar facets with solutions of different KOH content, requiring etching times as short as $130\,$s, and temperatures down to $40^{\circ}$C. 
%\hl{discuss diameter? }

%=====================================================================
%%Fig.2
\begin{figure*}
	\includegraphics*[width=.8\textwidth]{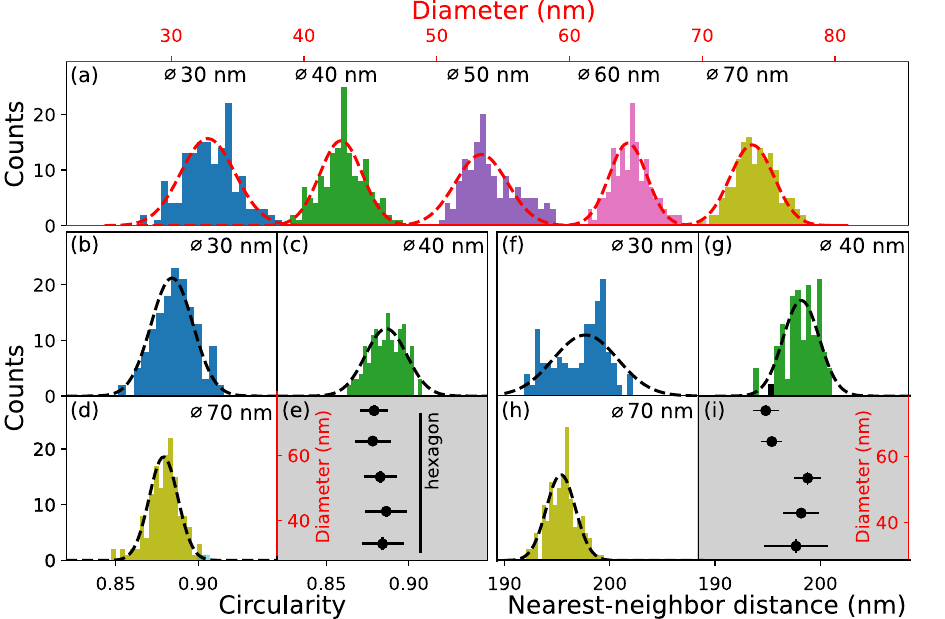}
	\caption{\label{02_uniformity}Statistical analysis of (a) the NW equivalent disk diameter, (b)--(d) the NW circularity, and (f)--(h) the NW nearest-neighbor distance taken for microfields with a nominal pitch of $200$\,nm. The nominal diameter is mentioned in each panel. The dashed lines are fits of the histograms by Gaussians. Panels (e) and (i) summarize for each microfields the average NW circularity and nearest-neighbor distance as a function of the nanowire diameter.}
%%SAS sample: MIO_GaN#6F1 (M91498)
%%ICP-KOH sample: MIO_GaN#6A1
\end{figure*} 

%NW morphology: statistics 

To assess the uniformity of the fabricated GaN NW arrays, we use ImageJ to analyze the top-facet morphology and position of $10^3$ NWs per microfield which are pictured top-view by scanning electron microscopy (SEM). The resulting distribution of NW equivalent disk diameter \footnote{$ d = \sqrt{\frac{4 A}{\pi}}$, with $A$ the NW top facet area}, circularity \footnote{$c = 4\pi AP^{-2}$, with $P$ the NW top facet perimeter}, and (center-to-center) nearest-neighbor distance are shown in Fig.\,\ref{02_uniformity}. 
The distributions of NW diameter peak close to the nominal values and exhibit standard deviations of $3$--$5$\,nm. Independently of the NW diameter, the circularity of the NW top facet amounts to $0.88$--$0.89$, which is close to that of a regular hexagon ($0.91$). Likewise, the nearest-neighbor distance approaches the nominal pitch value with a standard deviation of $2$--$7$\,nm. Thus, the uniformity of the arrays is very high and is clearly mostly limited by the patterning precision. The thinnest NWs that could be achieved with high yield are characterized by an average diameter of $(33 \pm 5)\,$nm.

% The investigated technique thus allows the fabrication of highly ordered and uniform GaN NW arrays with $99.9$\,$\%$ yield, and the arrays of thinnest NWs are characterized by diameters averaging to $33\,$nm with a standard deviation of $5\,$nm. Such NW arrays could be reproduced on several samples.
% %Top-down fabricated, ordered arrays of vertical GaN NWs with a comparable diameter have been reported in Ref.\citenum{Jo_2015}.  
% %E:\PhD\PhD_project\Morphology_statistics\Plot_and_fit_results\All_Data_ready_for_figure
% Thinner averaged diameters in ordered GaN\cite{Kazanowska_2022,Shih_2022}, GaN/(In,Ga)N\cite{Chen_2006} and (Al,Ga)N\cite{Shih_2022} top-down NWs were only observed at the tip of strongly tapered NWs.
% %note that Teng_2016 may have thinner NWs (see figs 3 and 4 in which experimental data show down to 15nm). However, they don't show SEM images and don't discuss these very small diameters. The SEM image features NWs with 80 nm thickness. 
%=====================================================================

%Part II: can the processing steps be reduced? 
%=====================================================================
%%Fig.3
\begin{figure}
	\includegraphics[width=\columnwidth]{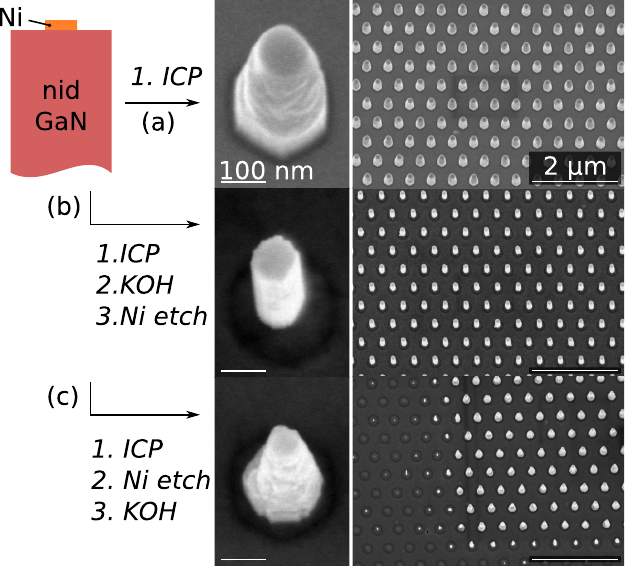}
	\caption{\label{03_roleSiN}Bird’s-eye view ($15\,^{\circ}$ tilt from top-view) secondary electron micrographs of GaN NWs after several etching steps. The initial mask is free of SiN$_x$. (a) shows the NWs after ICP etching, (b) after KOH etching and removal of Ni, and (c) after Ni etch first and KOH etching at last.}
%%a: MIO_GaN#23D (there are also SEM images of MIO_GaN#23A acquired after ICP) 
%%b: MIO_GaN#23B (other samples fabricated the same way: MIO_GaN#11C, MIO_GaN#11H, MIO_GaN#23B, MIO_GaN#23D)
%%c: MIO_GaN#23A (other samples fabricated the same way: MIO_GaN#11A, MIO_GaN#11D, MIO_GaN#11E, MIO_GaN#11G, MIO_GaN#23C)
\end{figure}

%Motivation
To address the benefits of the SiN$_x$ patch in the fabrication process, we further examine GaN NW arrays produced with a hard mask made of Ni only. 
% In the following, we discuss whether the top-down fabrication steps can be reduced. Due to the importance of Ni in the ICP etching mask mentioned before, the combination of a positive resist and a Ni layer can not be exchanged by a negative resist. However, the Si$_x$N$_y$ layer may not be needed for top-down fabrication of GaN NWs with ICP-KOH, and using a unique Ni layer would also lead to long and straight GaN NWs with smooth and non polar facets, and diameters down to $100\,$\cite{Liu_2015} and $140\,$\cite{Doundoulakis_2019}nm. 
% %Such a change would imply a simplification of the fabrication process, as the Si$_x$N$_y$ deposition and RIE etching steps could be omitted.
%
% discuss morphology after ICP
Figure\,\ref{03_roleSiN}(a) depicts that, upon ICP etching for $3\,$min, tapered GaN NWs exhibiting rough sidewalls and an hexagonal base are obtained, which compares well with the case where a Ni/SiN$_x$ mask is used (not shown here). 
%(NWs in Fig. \ref{01_synthesis} after the ICP step, not shown here)
This morphology is also consistent with previous studies \cite{Liu_2015,Doundoulakis_2019,Kazanowska_2022}. Straight NW sidewalls are eventually obtained during KOH etching [Fig.\,\ref{03_roleSiN}(b)], in agreement with previous reports \cite{Li_2011,Li_2012}. However, the roughness of the sidewalls is larger than in the presence of SiN$_x$. Remarkably, if the KOH etching is performed after removal of the Ni, the GaN NWs remain tapered and are eventually etched away, as seen in Fig.\,\ref{03_roleSiN}(c). The NWs first disappear at the edge of the microfields, suggesting that the etching rate is kinetically limited by mass transport in the KOH solution. The GaN (0001) facet at the NW tip is known to be resistant against KOH, whereas nonpolar and semipolar facets can be attacked by hot KOH \cite{Stocker_1998}. Here, our observations indicate that etching of thin GaN NWs in hot KOH can be mitigated by the presence of a SiN$_x$ cap or, to some extent, by a Ni cap placed at the NW tip. This finding is in agreement with the fact that the GaN NWs feature the same diameter as the SiN$_x$ cap [Fig.\,\ref{01_synthesis}(d)] and that the final NW diameter is found to be independent of the KOH etching conditions. 
As a result, it can be deduced that SiN$_x$ and Ni caps act as an etch stop by protecting the edge of the GaN (0001) top facet where the KOH attack otherwise proceeds. Such an etch stop is of high practical relevance since the final NW diameter is now solely determined by the size of the top patch written by EBL, hence providing higher reproducibility. 
The discovery that Ga-polar GaN NWs can be etched out by KOH is also of relevance if using KOH etching as a mean to determine the polarity of GaN NWs \cite{Zuniga-Perez_2016}. 

\subsection{Digital etching of the NWs}
%Part III: thinning down NWs

%=====================================================================
%%Fig.4
\begin{figure*}
	\includegraphics*[width=\textwidth]{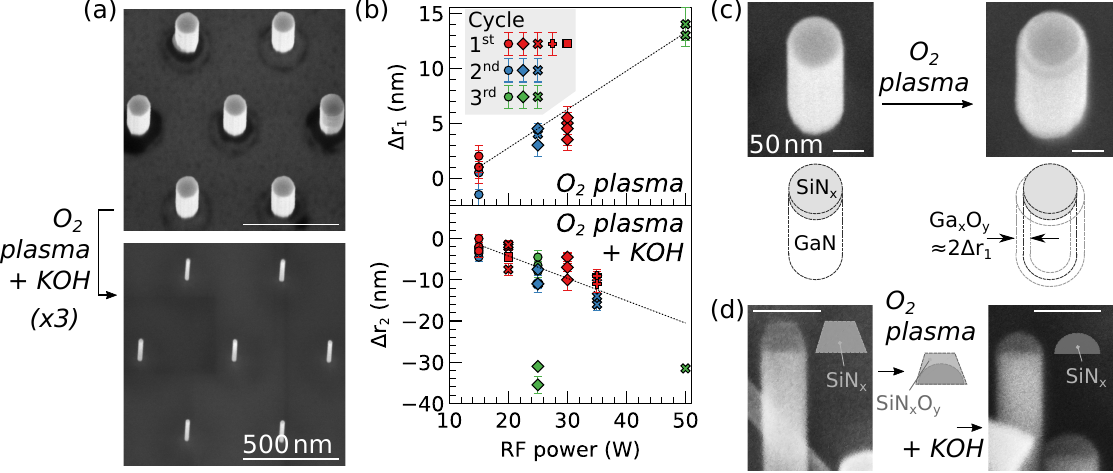}
	\caption{\label{04_thinning}(a) Bird's eye view ($20\,^{\circ}$ tilt from top-view) secondary electron micrographs of the same GaN NW field before and after several cycles of digital etching. (b) Relative change of the NW radius after O$_2$ plasma only and O$_2$ plasma + KOH etching as function of the RF power during O$_2$ plasma. Symbols refer to different samples (including microfields of different pitch/diameters) and colors refer to the number of treatments already applied on the same sample. (c)--(d) Magnified view of single NWs from the same NW field before and after exposure to (c) a $50$\,W O$_2$ plasma and to (d) a $20$\,W O$_2$ plasma + KOH etching. Panel (c) additionally depicts a schematic view of the expected NW configuration before and after the O$_2$ plasma treatment based on Ref.\,\cite{Bui_2019}. In (d), some NWs were accidentally broken, which allows imaging of the SiN$_x$ cap. Here, a schematic view shows the morphology of the SiN$_x$ patch observed after each step, and its composition, assuming that KOH etches the SiN$_x$O$_y$ layer created during the O$_2$ plasma treatment. The scale bars in (c) and (d) all amount to $50\,$nm. }
%% (a): MIO_GaN15B, microfield 120_500, before thinning (left), after 3 cycles of thinning procedure (right)
%% (b): MIO_GaN12A,B, _GaN15A,B, _GaN21E,F,G
%% (c): MIO_GaN15B, microfield 120_500, before thinning (left), after O2 etching (right)
%% (d):	MIO_GaN15A, microfield 40_200, before thinning (left), and after 2 cycles of thinning procedure (rigth)
\end{figure*}

%Motivation/qualitative analysis (morphological)
% In the approach described so far, the NW diameter is determined by the size of the initial Ni/Si$_x$N$_y$ patch produced by EBL patterning and RIE etching, and we achieved a

The minimum NW diameter achieved here by direct selective area etching amounts to ($33\pm5$)\,nm, a value that is likely limited by the precision of the EBL patterning of the Ni/SiN$_x$ hard mask. We thus explore digital etching to further decrease the NW diameter. The digital etching approach consists in plasma oxidation of the NWs followed by etching in hot KOH. This process is shown to be effective in Fig.\,\ref{04_thinning}(a), where NWs initially $120$\,nm thick are substantially reduced in diameter by repeating three times the etching process with a RF power of $20$, $35$, and $50$\,W, respectively. The etching rate appears to be uniform along the NW length and similar for all NWs of the same sample. 

%Additionally, we observe that all GaN NWs still feature similar diameters as their Si$_x$N$_y$ top patches after the thinning procedure.

%notes taken during the discussion with Thomas, 08/07/2022. 
%there are 2 concurring processes for the thinning down: SixNy width reduction vs removal of GaxOy
%KOH doesn't etch GaxOy in general, but we can accept (maybe search lit for that) that KOH etches defective material (as well as subsoichiometric material). 
%the SixNy path is also expected to be reduced by the thinning procedure after each cycle. 
%the data points in 4(b) not following the linear behaviour show that it is not possible to have an infinite amount of cycles (each cycle depend on the RIE power)

%quantification of O2 etching
The etching rate as a function of the RF power is quantified by analysing SEM images of $20$--$50$ NWs before and after treatment, and the results are shown in Fig.\,\ref{04_thinning}(b). First, the O$_2$ plasma exposure actually increases the NW radius by an amount $\Delta r_{1}$ that scales linearly with the RF power. As seen in Fig.\,\ref{04_thinning}(c), only the GaN NW expands whereas the SiN$_x$ cap remains visually unaffected. A similar increase in the NW radius was reported in Ref.\,\cite{Bui_2019} for GaN NWs annealed in O$_2$ atmosphere and is associated to the formation of a GaO$_x$ shell with a width equal to $\approx 2\Delta r_1$. The size increase relates to the $18$\,\% larger specific volume of $\beta$-Ga$_2$O$_3$ compared to GaN (for an equal number of Ga atoms), to the epitaxial strain at the GaO$_x$/GaN interface, and to the abundance of defects. 
%If this mechanism applies for our NWs,  meaning that at a $\Delta r_{1} /2$ distance from the original GaN NW sidewall,
%\hl{does this mean that $\Delta r_1$ has not been measured but estimated based on the reference?}. 

%quantify KOH etching
After exposure to both O$_2$ plasma and hot KOH, the NW radius experiences a net decrease $|\Delta r_2|$, that scales linearly with the RF power. The linearity is lost when processing the NWs for the third time. Since $|\Delta r_2| > \Delta r_1$, we conclude that the NW thinning during KOH treatment is not limited to the removal of the GaO$_x$ shell formed during O$_2$ plasma. Instead, we propose that the NW thinning is limited by the radial etching of the SiN$_x$ cap. This is substantiated by Fig.\,\ref{04_thinning}(d), where the SiN$_x$ cap is seen to undergo a major change in morphology after etching and keeps sharing the same diameter as the underlying GaN stem. SiN$_x$ etching is further confirmed by treating SiN$_x$ layers deposited on GaN. X-ray reflectivity analysis on a planar reference sample (not shown here) reveals that O$_2$ plasma at a RF power of $40$\,W induces a pronounced chemical change at the layer surface and that wet etching in hot KOH eventually decreases the SiN$_x$ layer by $4$--$5$\,nm. However, this axial etching rate is three times smaller than the radial etching rate deduced for the SiN$_x$ cap, an effect attributed to the anisotropy of the SiN$_x$ layer itself and/or to the anisotropy of the O$_2$ plasma when applied on submicrometer high-aspect-ratio structures \cite{Kruger_2019}. We eventually associate the nonlinearity in the NW thinning rate after $3$ cycles of digital etching to the concurrent effect of radial and axial etching of the SiN$_x$ cap. 
%planar SiNx sample: DSA_GaN15t3

\subsection{Mechanical instability of ultrathin NWs}

\begin{figure*}
	\includegraphics[clip,width=1\textwidth]{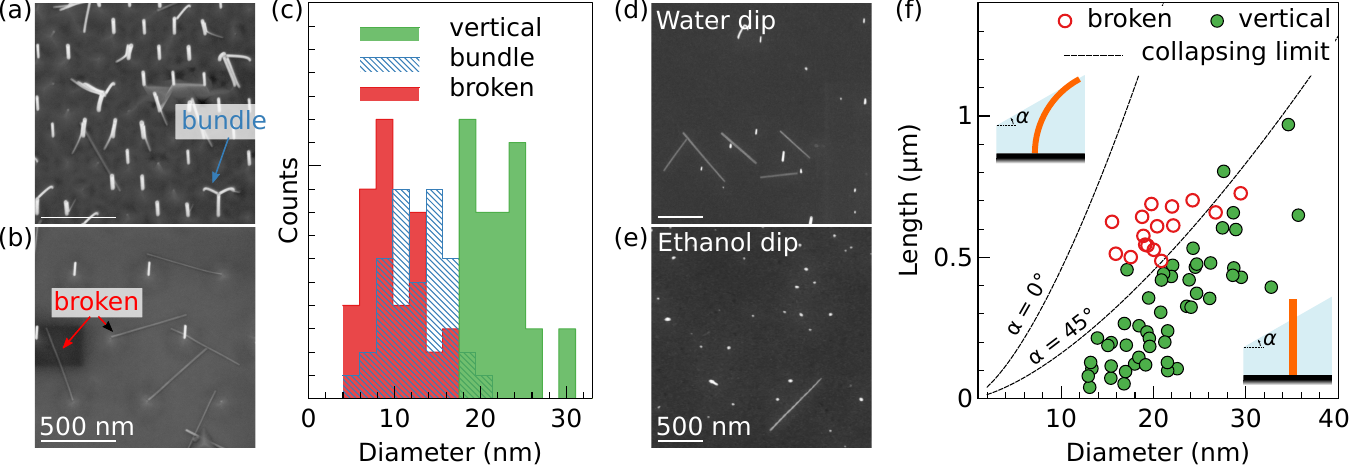} \caption{\label{05_collapsing}(a)--(b) Exemplary bird's eye view ($20\,^{\circ}$ tilt from top-view) secondary electron micrographs of ultrathin GaN NW microfields where digital etching has resulted in collapsing of most NWs. The NWs are $0.55$\,$\mu$m long and the NW pitch in (a) and (b) is $200$ and $500$\,nm, respectively. (c) Diameter histogram of the NWs shown in (a) that remained either vertical, that have bundled or that broke after the process. (d)--(e) Exemplary bird's eye view ($20\,^{\circ}$ tilt from top-view) secondary electron micrographs of thin self-assembled GaN NWs grown on TiN that were dipped in water and ethanol, respectively, and dried in air. (f) Plot of length versus diameter for broken and vertical NWs after water drying in air. Dashed lines correspond to the NW collapsing limit calculated for different water surface angles $\alpha$. During drying, strong bending leading to NW breaking is only expected above the collapsing limit, as schematized in the insets.}
\end{figure*}

NW diameters down to $5$\,nm are achieved using the thinning procedure.
However, these ultrathin NWs tend to collapse by lateral bundling or breaking, as exemplified in Fig.\,\ref{05_collapsing}(a) and Fig.\,\ref{05_collapsing}(b). Broken NWs crack at their anchor point with the substrate and lie nearby whereas bundled NWs touch each other at their tip and remain bent as a result of van der Waals or electrostatic forces. We note that broken and bundled NWs are observed already from the first scans of SEM imaging and, thus, cannot be attributed to electrostatic charging occurring during electron irradiation \cite{Loitsch_2015,Zettler_2016}. The histogram in Fig.\,\ref{05_collapsing}(c) shows the diameter distribution of $0.55$\,$\mu$m long NWs taken from the same field that remained either vertical, suffered from bundling or broke during the thinning process. It evidences the existence of two critical diameters amounting to $22\pm2$ and $17\pm2$\,nm,
below which NWs bundle and break, respectively.
%It evidences the existence of two critical diameters amounting to $22\pm2$ and $17\pm2$\,nm, below which NWs bundle and break, respectively. \hl{How did you determine these critical diameters? There are no vertical NWs below 18 nm, the thickest bundled NWs have a diameter of 21 nm, the thickest broken NWs of 17 nm, and the thinnest bundled NWs have a diameter of 5 nm. Clearly, there are bundled NWs that are much thinner than the critical diameter of 17 nm for breaking.}

We associate this mechanical instability to capillary forces acting on the NWs during the drying step. These forces become non-negligible for nano- and microstructures with increasingly high aspect ratio, having a major impact on their mechanical stability \citep{Chandra_2009,Tang_2010,Roman_2010,Togonal_2014,Park_2016,Neukirch_2007,Mallavarapu_2020,Han_2021}.
To test for this hypothesis, we subsequently wet and dry in air ensembles of thin GaN NWs obtained bottom-up by self-assembly on a TiN substrate. %{reference to your paper?}. The paper is not published yet, but yes if it is then we should.
Such ensembles feature a broad NW length distribution, which allows determining the critical NW dimensions possibly leading to failure. The micrographs in Figs.\,\ref{05_collapsing}(d) and \ref{05_collapsing}(e) confirm that the sole wetting and drying is indeed sufficient to break some NWs, with a more pronounced effect when using water as solvent compared to ethanol. This difference can be explained by the $3.5$ times larger surface tension of water compared to ethanol. Figure\,\ref{05_collapsing}(f) presents the lengths and diameters of NWs that have either collapsed or remained vertical after the water dip and drying. The plot clearly reveals two regions: the longer and thinner NWs break, while the shorter and thicker ones remain vertical.

%=====================================================================
%%Fig.5

%=====================================================================

Neukirch \textit{et al.} \citep{Neukirch_2007} have described the various buckling configurations that can be reached for a vertical elastic rod piercing a horizontal liquid surface. The NW buckling limit deduced from this work and plotted in Fig.\,\ref{05_collapsing}(f) as a dashed line labelled $\alpha = 0^\circ{}$ is seen not to match our experimental observations. As detailed below, an excellent agreement is found instead when considering an inclined liquid surface ($\alpha \neq 0^\circ{}$), as can be expected when the solvent has a finite wettabibity on the substrate surface and forms droplets.

% in detail a large variety of buckling states occurring for an elastic rod under the action of capillary forces. These states are  We show below that the buckling under the horizontal
% surface of a water layer does not explain our observations. Then,
% we consider the buckling under a water droplet and show that it can
% explain the breaking of NWs revealed in Fig.\,\ref{05_collapsing}(f).

\begin{figure}
	\includegraphics[width=1\columnwidth]{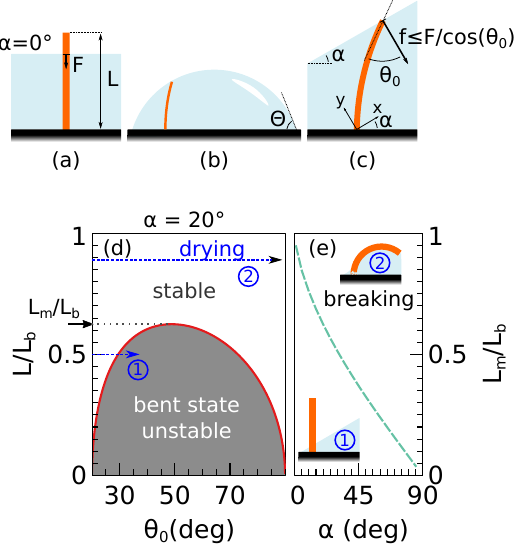} \caption{\label{06_model} (a) Case of a vertical wire piercing the horizontal surface of a solvent. (b) Case of a NW embedded in a droplet. (c) Case of a bent wire about to pierce the inclined surface of a solvent. (d) Stability diagram of the GaN NW bent state in water under the capillary force $f\leq F/\cos\theta_0$ calculated for $\alpha = 20 ^\circ{}$. The blue dotted arrows depict two different bending paths occurring during water drying, whether the NW length exceeds $L_m$ or not. (e) Critical length $L_{m}/L_{b}$ as function of $\alpha$. For NWs with a length exceeding $L_{m}$, breaking is expected after complete drying of the solvent, as sketched in the inset.}
\end{figure}

Let us consider a vertical NW of length $L$ placed in water, under the horizontal liquid surface. By lowering the water level, the liquid surface comes in contact with the NW tip [Fig.\,\ref{06_model}(a)] and exerts a vertical force at the NW tip amounting to
\begin{equation}
	F=2\pi r\gamma\cos\Theta,\label{eq:1}
\end{equation}
with $r$ the NW radius, $\gamma$ the solvent surface tension, and $\Theta$ the water wetting angle at the NW sidewalls. According to the Euler criterion \citep{Neukirch_2007,Landau_1986}, the NW will eventually buckle if its length exceeds

\begin{equation}
	L_{b}=(\pi/2)\sqrt{EI/F},\label{eq:2}
\end{equation}
with $E$ the NW Young modulus, and $I=\pi r^{4}/4$ the geometrical moment of inertia of the NW cross section. Taking the water surface tension $\gamma=72$\,mN/m, the water wetting angle on GaN $\Theta=40^{\circ}$ \citep{Dziecielewski_2013}, and the GaN Young modulus $E=300$\,GPa \citep{Bernal_2011}, we obtain $L_{b}\approx 40r^{3/2}$, with $L_{b}$ and $r$ expressed in nanometers. It is this buckling limit for a horizontal liquid surface ($\alpha = 0^\circ{}$) that is plotted in Fig.\,\ref{05_collapsing}(f) as a dashed line. This limit is seen to largely overestimate the threshold at which NW breaking is experimentally observed. A lower threshold would require either a reduced mechanical strength of the NWs, or a higher surface tension of the solvent, both of which are not physically reasonable assumptions.

%=====================================================================

Compared to the case described in Ref.\,\cite{Neukirch_2007}, we deal here with much shorter wires, which can make the assumption of a horizontal liquid surface not valid anymore. Since water does not wet either TiN or GaN substrate surfaces \cite{Dziecielewski_2013,Colovic_2019}, it is reasonable to assume that water has already assembled in droplets when the liquid surface comes in contact with the NW tip [see Fig.\,\ref{06_model}(b)]. The inclination angle $\alpha$ of the liquid surface at the NW tip may thus take any value between $0^{\circ}$ and $\Theta$, with $\Theta$ the water wetting angle on the NW substrate surface. The direction of the capillary force is normal to the liquid surface and is thus misaligned with the NW axis for $\alpha>0^\circ{}$, which will always result in some bending of the NWs. The capillary force is proportional to the perimeter of the NW cross-section in the plane of the liquid surface \cite{Neukirch_2007}. Compared to the case $\alpha = 0^\circ{}$, the amplitude of the capillary force thus increases by a factor  $1/\cos\theta_{0}$, with $\theta_0$ the angle between the NW axis at its tip and the normal to the liquid surface [see Fig.\,\ref{06_model}(c)]. 

During drying, once the liquid surface touches the NW tip and goes lower, the NW starts bending ($\theta_0$ increases), which raises in turn the strength of the capillary force $F/\cos\theta_{0}$ eventually resulting in some more bending. In this process, $\theta_0$ evolves from $\alpha$ (no deflection) possibly up to $90^\circ{}$ (NW tip axis parallel to the liquid surface). In the course of bending, however, the strength of elastic restoring force may exceed the capillary force, which will result in the NW piercing the liquid surface and getting back to its vertical configuration ($\theta_0 = \alpha$). 

Calculations of stable NW bending configurations as function of $\alpha$ are detailed in the Appendix and the results are presented in Fig.\,\ref{06_model}(d) for $\alpha=20^{\circ}$. NW lengths are given in units of the buckling length $L_{b}$ and $\theta_0$ ranges from $\alpha$ to $90^\circ{}$. For bending states falling in the grey area, the elastic restoring force is larger than the capillary force, making these states unstable. The maximum NW length within the unstable region is denoted by $L_m$.
During sample drying, NWs follow two different paths, depending on their length exceeding $L_m$ or not. These paths are exemplified by the two blue dotted arrows in the stability diagram. For path $1$ ($L<L_m$), $\theta_0$ reaches its maximum value at the boundary of the unstable area. If the water level further decreases, the NW will pierce the water surface and straighten up, as shown in the inset of Fig.\,\ref{06_model}(e). In contrast, for path 2 ($L>L_m$), $\theta_0$ can reach the maximum value of $90^\circ{}$. As detailed in the Appendix, we have checked that the bending states that can be achieved for $L<L_m$ do not build a sufficient stress for inducing NW breaking. In contrast, we propose that once $\theta_0$ reaches $90^\circ{}$ the NW will not be released from the droplet as it shrinks in size and the NW will thereby enter complex bending configurations resulting in breaking. The breaking condition thus becomes $L>L_m$. Figure\,\ref{06_model}(e) shows that $L_{m}$ monotonously decreases from $L_{b}$ to 0 when $\alpha$ increases from $0$ to $90^\circ{}$.

As can be seen in Fig.\,\ref{05_collapsing}(f), an excellent agreement between calculations and the experimental threshold at which GaN NWs on TiN are breaking can be obtained by taking $\alpha=45^{\circ}$ ( $L_{m}/L_{b} = 0.37$). Similarly, $\alpha=25^{\circ}$ ($L_{m}/L_{b}=0.55$) can account for the observed breaking of GaN NWs on GaN [Fig.\,\ref{05_collapsing}(c)]. These values are reasonably close to reported water wetting angles on TiN ($32^\circ{}$ \cite{Colovic_2019}) and GaN ($40^\circ{}$ \cite{Dziecielewski_2013}), taking into account the fact that the droplet itself is likely deformed in the presence of many NWs. We thus conclude that NW breaking is initiated in the vicinity of droplet edges. However, the bent NWs may escape their fate if their tip ends up touching neighboring NWs in the droplet, eventually forming tripod structures as can be seen in Fig.\,\ref{05_collapsing}(a). These bundled NWs are mechanically more robust than single NWs and may not break.

Several options can be used to prevent the collapsing of ultrathin NWs occurring during drying. One can either work with shorter NWs ($L<L_m$), use a critical point dryer as typically used during fabrication of microelectromechanical systems, or perform dedicated surface treatments as described in Ref. \cite{Ghosh_2022}.

\section{Conclusions}
%We have introduced here a facile route for producing ordered arrays of down to $5\,$nm-ultrathin GaN NWs with an aspect ratio exceeding $10$. Thin GaN NWs  
%(diameter $<120$\,nm) 
We have introduced here a facile route for the top-down fabrication of ordered arrays of GaN NWs with aspect ratios exceeding $10$ and diameters below $20\,$nm. Thin NWs are first obtained top-down by EBL patterning of a Ni/SiN$_x$ hard mask followed by dry etching and wet etching in hot KOH. SiN$_x$ in the hard mask is found to work as an etch stop during wet etching in hot KOH. Without SiN$_x$ caps, our observations show that Ga-polar GaN NWs can be completely etched out in hot KOH. The fabricated NW arrays show excellent uniformity and $99.9\,\%$ yields for NW diameters down to $(33 \pm5)$\,nm. Further reduction of the NW diameter is obtained by applying digital etching which consists in cycles of plasma oxidation and wet etching in hot KOH. The radial etching depth is limited by the SiN$_x$ and can be tuned by varying the RF plasma power during plasma oxidation. The digital etching allows reducing NW diameters down to $5$\,nm. However, NW breaking or bundling is observed when the diameter falls below $\approx 20$\,nm, an effect that is associated to capillary forces acting on the NWs during sample drying in air. The mechanical failure of NWs is found to occur at much smaller aspect ratio than predicted for models dealing with macroscopic elastic rods. This discrepancy is further associated to the fact that water eventually forms droplets during drying, which implies that the liquid surface is not horizontal. Explicit calculations taking into account an inclined liquid surface deliver an excellent agreement with the breaking threshold determined experimentally. To prevent collapsing of high-aspect ratio NWs by capillary forces, drying should be performed in a critical point dryer. 

The proposed thinning approach can be principally applied to GaN/SiN$_{x}$ nanostructures of arbitrary shape and allows regrowth after removal of the SiN$_{x}$ mask. The ultrathin nanostructures produced here can be used as template for epitaxy of highly mismatched materials \cite{balaghi_2019}, or as channel in ballistic deflection transistors \cite{Chowdhury_2017}. 

\section{Acknowledgements}
The authors acknowledge valuable experimental support by Jonas Lähnemann 
%and Timur Flissikowski, 
and technical support by Carsten Stemmler, Katrin Morgenroth, Anne-Kathrin Bluhm, Walid Anders, Sander Rauwerdink, Werner Seidel and Nicole Bickel, and Philipp John for a critical reading of the manuscript. %\hl{add Jonas as an author?in icpkoh vs SAS maybe}

\section*{Appendix: NW bending under an inclined liquid surface}

We consider here a thin elastic rod with circular cross-section that is referred to as a NW. The NW is placed in a liquid having a surface inclined by an angle $\alpha$ with respect to the horizontal. The height of the meniscus that forms once the NW pierces the liquid surface is on the order of the NW radius \citep{Derjaguin_1946,James_1974} and is thus further neglected. We emphasize that for macroscopic rods as described by Neukirch \emph{et al.} \citep{Neukirch_2007}, the meniscus height scales instead with the capillary length $L_c = \sqrt{\gamma/(\rho g)}$, with $\gamma$ the surface tension, $\rho$ the liquid density, and $g$ the gravitational acceleration ($L_c \approx 2$\,mm for water).

When the NW comes in contact with the liquid surface, a capillary force $f$ acting on the NW tip builds up and points in a direction normal to the liquid surface [see Fig.\,\ref{06_model}(c)]. Its component tangential to the liquid surface is zero since a displacement along the liquid surface does not cost energy. The amplitude of $f$ is proportional to the perimeter of the NW cross-section in the plane of the liquid surface. We denote by $\theta_{0}$ the angle between the NW axis at its tip and the normal to the liquid surface [see Fig.\,\ref{06_model}(c)]. For fixed $\theta_0$, $f$ reaches its maximum value $F/\cos\theta_0$, with $F$ given by Eq.\,(\ref{eq:1}), once the NW completely pierces the liquid surface \cite{Neukirch_2007}. After piercing, the elastic restoring force becomes larger than the capillary force which makes the bent state unstable and leads to a straightening of the NW \cite{Neukirch_2007}.

Our aim is to find the stability diagram of the NW bent states under the action of capillary forces. To this end, we calculate the NW equilibrium state assuming the maximum capillary force $F/\cos\theta_0$, which corresponds to the border of the stable region. The NW bending configuration can be described using the general solution of Problem 1 to §19 in Ref.\,\citep{Landau_1986}. Let us take the plane of bending as the $xy$ plane and direct the $y$ axis antiparallel to the capillary force acting at the NW tip [see Fig.\,\ref{06_model}(c)]. The $x$ axis is parallel to the liquid surface and makes an angle
$\alpha$ to the substrate. Let $\theta(l)$ be an angle between the $y$ axis and the NW axis at a distance $l$ along the NW length ($l=0$ at the NW anchor point on the substrate). Then, the NW shape is described by the equation \citep{Landau_1986}
\begin{equation}
	\frac{IE}{2}\left(\frac{d\theta}{dl}\right)^{2}-f\cos\theta=c,\label{eq:4}
\end{equation}
where the constant $c$ has to be determined from boundary conditions. We have changed the sign of the second term in Eq.\,(\ref{eq:4}) in comparison to Ref. \citep{Landau_1986} since the direction of the force $f$ is opposite to the direction of the $y$ axis. The bottom of the NW is clamped normal to the substrate, so that $\left.\theta\right|_{l=0}=\alpha$ in our coordinates. The NW tip is free from a force moment resulting in $\left.d\theta/dl\right|_{l=L}=0$. We have already denoted $\left.\theta\right|_{l=L}=\theta_{0}$. Then, $c=-f\cos\theta_{0}$ and the solution of Eq.\,(\ref{eq:4}) is
\begin{equation}
	l(\theta)=\sqrt{\frac{EI}{2f}}\int_{\alpha}^{\theta}\frac{d\theta}{\sqrt{\cos\theta-\cos\theta_{0}}}.\label{eq:5}
\end{equation}
Substituting $f=F/\cos\theta_{0}$ and using the definition of the buckling length $L_{b}$ in Eq.\,(\ref{eq:2}) results in a bent state 
\begin{equation}
	\frac{L}{L_{b}}=\frac{\sqrt{2\cos\theta_{0}}}{\pi}\int_{\alpha}^{\theta_{0}}\frac{d\theta}{\sqrt{\cos\theta-\cos\theta_{0}}}.\label{eq:6}
\end{equation}

The line defined by Eq.\,(\ref{eq:6}) is plotted in Fig.\,\ref{06_model}(d) in the case of a GaN NW placed in water with $\alpha=20^{\circ}$. The bending configurations below the line are unstable since one would need a capillary force exceeding $F/\cos\theta_0$ to reach such states. During drying, $\theta_0$ increases, following the paths exemplified by blue arrows in Fig.\,\ref{06_model}(d). For $L<L_m$, the bending path enters the unstable region, meaning that the NW will eventually pierce the water surface. In contrast, for $L>L_m$, the NW will always remain below the water surface, likely reaching complex bending configurations as the droplet shrinks in size. Their calculation is beyond of the theoretical framework presented here. The bent states located on the right side of the unstable area cannot be reached during drying. 

It remains to check, if the elastic stress building up in bent NWs with $L\leq L_m$ is sufficient to induce NW breaking during drying. During bending, there is a critical surface stress $\sigma_{c}$ above which NWs will crack. The longitudinal stress at the NW surface depends on the local radius of curvature $R$ and can be expressed as $\sigma=Er/R$. Hence, the breaking criterion can be written as
\begin{equation}
	\sigma_{c}<E\frac{r}{R}.\label{eq:10}
\end{equation}

For a fragile material, a critical stress $\sigma_{c}=2\gamma/a$ can be estimated as a force needed to break chemical bonds and separate from each others the two newly formed surfaces by a distance equal to the bond length $a$ (see, \emph{e.g.}, Ref.\,\citep{Friedel_1964}). For a GaN surface energy $\gamma_{\mathrm{GaN}}$ of approximately $100$\,meV/$\textrm{Å\ensuremath{^{2}}=1.6}$\,N/m$^{2}$ \citep{Northrup_1996,Dreyer_2014,Li_2015} and $a=3.2$ Å, we get $\sigma_{c}=10$
GPa, fairly close to values reported for CuO and ZnO NWs broken by AFM tips \cite{Polyakov_2012}. 

\begin{figure}
	\includegraphics[width=1\columnwidth]{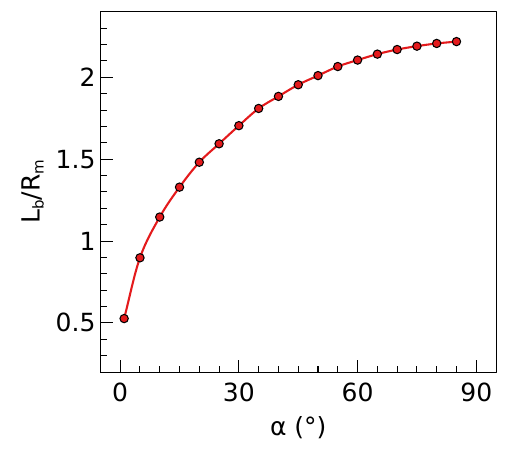}
	
	\caption{Maximum curvature accessible for NW length $L=L_m$ during drying as function of the liquid surface angle $\alpha$.}
	
	\label{fig:Rmin}
\end{figure}

The NW curvature along its length $d\theta/dl$ is maximum at the NW bottom and can be obtained from Eq.\,(\ref{eq:4}). Using again the definition of $L_{b}$ from Eq.\,(\ref{eq:2}), we find for $1/R=\left.d\theta/dl\right|_{l=0}$
\begin{equation}
	\frac{L_{b}}{R}=\pi\left(\frac{\cos\alpha-\cos\theta_{0}}{2\cos\theta_{0}}\right)^{1/2}.\label{eq:7}
\end{equation}
The right-hand side of Eq.\,(\ref{eq:7}) is monotonically increasing with $\theta_{0}$. Hence, for $L\leq L_m$ and for bent states accessible during drying, the minimum bending radius $R_m$ is obtained at the boundary of the unstable region for $L = L_m$. Fig.\,\ref{fig:Rmin} shows the dependence of $L_{b}/R_{m}$ on the liquid surface angle $\alpha$. In the range $\alpha<45^{\circ}$ that is of interest here, we have $L_{b}/R_{m}<2$. Substituting this maximum curvature in Eq.\,(\ref{eq:10}) and using a buckling length of $L_{b}=40r^{3/2}$ as mentioned in the main text for GaN NWs in water, we obtain that only NWs with $r<2$\,nm would crack. 
Such small radius is out of the experimental range, meaning that NW breaking for $L<L_m$ can be neglected. The main breaking condition thus remains $L>L_{m}$, as used in the main text.

%=====================================================================
\bibliographystyle{iopart-num}

\bibliography{Top_downGaNNWs_ICP-KOH}

\end{document}